\documentclass[12pt]{spieman}  
\usepackage{amsmath,amsfonts,amssymb}
\usepackage{graphicx}
\usepackage{setspace}
\usepackage{tocloft}
\title{Measurement of Out-of-Plane first-order Displacement Derivatives in Orthogonal shear directions Using Dichroic Mirrors}

\author[a]{Yinhui Guo}
\author[a]{XinDa Zhou}
\author[a]{Jie Li}
\author[a]{Rongsheng Ba}
\author[a]{YinBo Zheng}
\author[a,*]{Liqun Chai}
\affil[a]{Laser Fusion Reasearch Center, China Academy of Engineering Physics, Mianyang 621900, China }

\cftpagenumbersoff{figure}
\cftpagenumbersoff{table} 
\begin{document} 
\maketitle

\begin{abstract}
This paper proposed a novel and temporal phase-shift digital shearography system for simultaneous measurement of first order displacement derivative in orthogonal shear directions. Dual lasers with wavelengths of 532nm and 637nm, three splitter prism structure, two dichroic mirrors with different response wavelength, and the color CMOS are used in the system. Two dichroic mirrors can be used as shear mirrors to realize shearing in orthogonal directions at the same time. The system realizes the measurement of the first-order displacement derivative information in the orthogonal direction of the round metal aluminum plate of diameter 250mm. The experimental results show that the overall displacement integral PV error in the x and y directions is 2.7\%-14.8\%, which verified the reliability of the system.
\end{abstract}


{\noindent \footnotesize\textbf{*Liqun Chai},  \linkable{chailiqun@163.com} }

\begin{spacing}{2}   

\section{Introduction}

Intrinsic defects, new defects or performance degradation of materials will cause stress concentration or mutation, which will affect its safety, reliability and working life, and will lead to major accidents in severe cases~\cite{H,O,S,Chen,W}. Nondestructive testing (NDT) of materials can detect the presence or absence of defects, evaluate material or product quality, and measure physical or mechanical properties, thereby controlling and improving quality and ensuring safe operation of equipment~\cite{A,Garnier}. NDT techniques currently in use include magnetic particle detection, eddy current testing, penetration testing, ultrasonic testing, x-ray detection, acoustic emission testing, infrared thermal wave detection, microwave detection and digital shearography testing ~\cite{K,D,Gholizadeh,Dwivedi,B}. Compared to other NDT techniques, digital shearography has the advantages of full-field, real-time, non-contact, high sensitivity and anti-environmental interference ~\cite{Hung,L}. Hence, it has become an important tool in insitu detection of defects such as aeronautical composite materials, metal samples, Automobile tires, etc.
Digital shearography is a laser-based optical interferometry technique, which can directly measure the first-order displacement derivative. The commonly used phase extraction methods in digital shearography are temporal phase-shift method and spatial phase-shift method~\cite{Q}. The temporal phase-shift method has low difficulty in implementation high measurement accuracy, large dynamic range and is suitable for static or quasistatic measurements, while the spatial phase-shift method is suitable for a continuous or dynamic loading measurement. 

Under external excitation, the stress concentration in the defects can be judged by the abrupt change of speckle interference fringes, and then identify the location and size of the defects. In digital shearography, the detection rate of defects is affected by loading mode and amount, shape and depth of defects, and shear direction and amount~\cite{L,Q,Jiang}. In digital shearography, when the shear amount is half of the minimum defect radius on the sample surface, the detection sensitivity of the system is the highest~\cite{L}. Digital shearography is insensitive to the strain caused by defects parallel to the shear direction, so the traditional system with a single shear direction leads to low defect detection rate. When digital shearography is applied to defect detection, the shear direction needs to be increased to improve the defect detection rate. A dual shear directional digital shearography system composed of two CCDs and two polarization beamsplitters for simultaneous measurement of the out of plane first-order displacement derivative equivalent to two shear directions has been depicted in reference ~\cite{Steinchen}. But two CCDs increase the complexity of the system and make it difficult to adjust. A Michelson interferometer with improved three-beam splitter structure is described in reference~\cite{Bai}. The first out-of-plane displacement derivative in orthogonal direction is measured by time-sharing and step-by-step measurement by changing the shear direction by two shutters. However, the synchronous measurement of the out-of-plane first-order displacement derivative in the orthogonal direction is not realized, which takes a long time, the stability is poor, and the field of view is small. Two Michelson Interferometers are used as the shear device for simultaneous dual directional strain measurement in reference~\cite{AYW}. A digital shearography system which combinates a dual-wavelength light source with the 3CCD and adds a dichroic mirror behind a Michelson mirror to realize shear in the other direction is proposed in reference~\cite{Jiang}. It realizes the synchronous measurement of the first derivative of out-of-plane displacement in the orthogonal direction, but the field of view is limited without combining 4f system. The system currently used in measuring the out-of-plane displacement first-order derivative in orthogonal shear directions is complex and difficult to adjust, and the field of view of the measurement is limited. 

To solve these problems, a novel temporal phase-shift digital shearography system combining dual lasers with wavelengths of 532nm and 637nm, two dichroic mirrors with different response wavelength, and the color CMOS is proposed. The system uses dual lasers with wavelengths of 532nm and 637nm to synchronize paraxial illuminating, and uses dichroic mirrors with different working wavelengths to replace shear mirrors, and realizes shear in orthogonal direction at the same time. The introduction of the three splitter prism structure extends the traditional Michelson interferometer to balance the light intensity and introduces a 4f system to expand the field of view. In this system, the color CMOS is used to record the speckle patterns, and the separation of the R and G channels signals is realized, in which the optimal response wavelengths of the R and G channels in the color CMOS match the emission wavelengths of the two lasers, respectively. The PVs of displacement along the x and y shear directions are consistent which verifies the reliability of the system. The proposed system can be applied to synchronous measurement of the out-of-plane first-order displacement derivatives in orthogonal shear directions and NDT of multi-directional defects.

\section{Theory}

\begin{figure}
\begin{center}
\begin{tabular}{c}
\includegraphics[width=14cm]{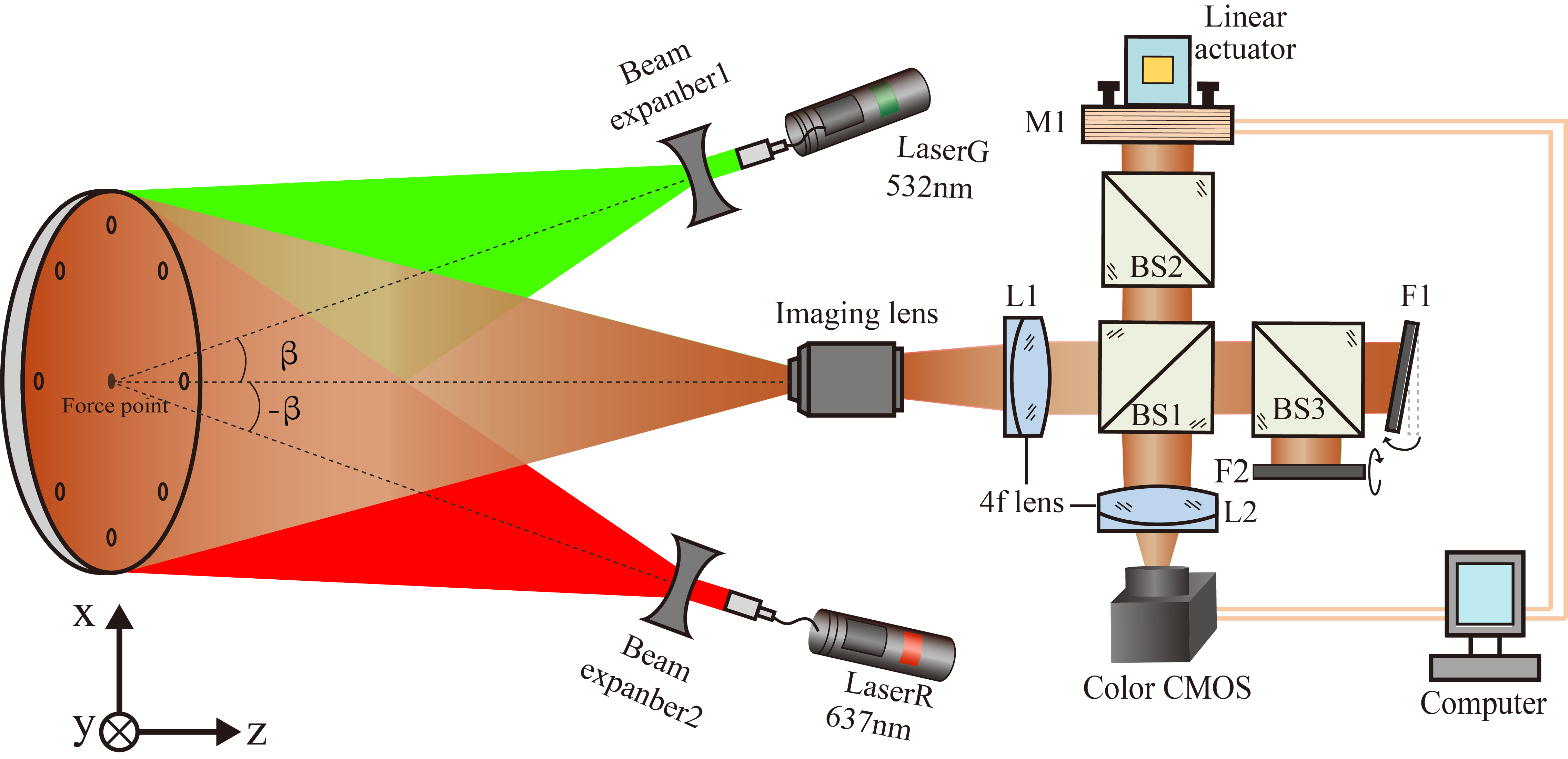}
\end{tabular}
\end{center}
\caption{Schematic diagram of the novel Michelson}
\label{f1}
\end{figure}

Schematic diagram of the novel Michelson Interferometer based temporal phase-shift shearographic system is illustrated in Figure~\ref{f1}. The system mainly includes: two single-mode fiber lasers (Laser R and Laser G) with different wavelengths ($\lambda_{R}=637 \mathrm{~nm}, \lambda_{G}=532 \mathrm{~nm}$), beam expanber1 and 2, imaging lens, two double-glued lens $L_{1}$ , $L_{1}$ with a focal length of 130mm, beam splitters $BS_{1}$, $BS_{2}$, and $BS_{3}$, the reflecting mirror $M_{1}$ with linear actuator, dichroic Mirror $F_{1}$ with transmission wavelength 532nm and reflection wavelength 637nm, dichroic Mirror $F_{1}$ with transmission wavelength 637nm and reflection wavelength 637nm, the color CMOS, computer for controlling lasers and linear actuator and processing shearography-interferometric phase fringe patterns, and the test object.

The three beam splitters ($BS_{1}$, $BS_{2}$, and $BS_{3}$) in the system are located on the three vertices of the isosceles right triangle, respectively. Two double-glued lens $L_{1}$, $L_{1}$ composes 4f system for image transmission. The total optical path of between the mirror $M_{1}$ and double glued lens $L_{1}$, double-glued lens $L_{2}$ is 260mm which is the sum of the focal lengths of double glued lens $L_{1}$ and double glued lens $L_{2}$. The total optical path of between dichroic mirror $F_{1}$ and double glued $L_{1}$, double glued lens $L_{2}$ is 260mm which is the sum of the focal lengths of double glued lens $L_{1}$ and double glued lens $L_{2}$. The total optical path of between dichroic mirror $F_{2}$ and double glued $L_{1}$, double glued lens $L_{2}$ is also 260mm which is the sum of the focal lengths of double glued lens $L_{1}$ and double glued lens $L_{2}$.

Two expanded lasers (Laser R and Laser G) are used to illuminate the test object paraxially. The diffuse reflection light formed enters beam splitters $BS_{1}$ through the imaging lens and double glued lens $L_{1}$, and is divided into the first reflected light and the first transmitted light. The first reflected light enters beam splitter $BS_{1}$, and is divided into the second reflected light and the second transmitted light. After the second transmitted light reaches the reflecting mirror $M_{1}$, it is reflected, then transmitted by beam splitter $BS_{2}$ and beam splitter $BS_{1}$. It is received by color CMOS through double glued lens $L_{2}$, where the received reference diagram is recorded as $T$. The first transmitted light enters beam splitter $BS_{3}$ and is divided into the third reflected light and the third transmitted light. The third transmitted light reaches dichroic mirror $F_{1}$, and the light of wavelength 532nm is transmitted and the light of wavelength 637nm is reflected. The reflected light of wavelength 633nm is transmitted by beam splitter $BS_{3}$ and reflected by beam splitter $BS_{1}$ It is received by color CMOS after through double glued lens $L_{2}$ where the received shearography patterns is recorded as $T_{1}$. The third reflected light reaches dichroic mirror $F_{2}$, and the light of wavelength 637nm is transmitted and the light of wavelength 532nm is reflected. Then the light of wavelength 532nm is reflected by beam splitters $BS_{3}$ and $BS_{1}$. It is received by color CMOS after through double glued  $L_{2}$ where the received shearography patterns is recorded as $T_{2}$.

Adjust dichroic mirror $F_{1}$ so that $T$ and $T_{1}$ are misplaced on color CMOS and define as the first shear direction. Adjust dichroic mirror $F_{2}$ so that $T$ and $T_{2}$ are misplaced on the color CMOS and define it as the second shear direction. The first shear direction is orthogonal to the second shear direction and is defined as the $X$ direction and Y direction of the coordinate system, respectively. 

The formulas are introduced for better understanding of the whole process. 

As shown in Figure~\ref{f1}, dual-wavelength light uses a linear actuator installed behind the mirror to drive the mirror to change the optical path to achieve phase shifting. The two wavelengths of light produce different phase shiftings. The steps of the system to measure the first derivative of out-of-plane displacement in the orthogonal direction are as follows. Carré algorithm has the advantage of not knowing the specific phase shifting~\cite{Carré,ZhangY}. This method in this paper is used to calculate the phase of each pixel in the shearography-interferometric phase fringe patterns formed by two different colors of laser before and after displacement.

(1) When the sample is measured before loading, the mirror is moved three times with equal step size by controlling the phase-shifting device, and the size of each step is $\bigtriangleup L$. The shifting-phase increment corresponding to each step is $2 \alpha (\alpha=2 \pi \bigtriangleup  L f /\lambda_{j}, j=R,G, \lambda_{R}=532 \mathrm{~nm},$\\$ \lambda_{G}=637 \mathrm{~nm})$. The intensity of the shear speckle interferogram before loading can be expressed as:

\begin{equation}
\label{eq1}
\left\{\begin{array}{l}
I_{1}\left(x, y\right)=I_{a}+I_{b} \cos [\phi(x, y)-3\alpha(x, y)]\\
I_{2}\left(x, y\right)=I_{a}+I_{b} \cos [\phi(x, y)-\alpha(x, y)] \\
I_{3}\left(x, y\right)=I_{a}+I_{b} \cos \left[\phi(x, y)+\alpha\left(x, y\right)\right] \\
I_{4}(x, y)=I_{a}+I_{b} \cos [\phi(x, y)+3 \alpha(x, y)]
\end{array}\right.\,,
\end{equation}

Where $I_{1}\left(x, y\right),I_{2}\left(x, y\right),I_{3}\left(x, y\right),I_{4}\left(x, y\right)$ are speckle interferogram for each moving equidistant distance, respectively. $I_{a},I_{b}$ are the intensities of the object and reference beams, respectively. $\phi(x, y)$ is the phase difference between the reference and the object beam. $-3 \alpha(x, y),-\alpha(x, y), $\\$\alpha(x, y),3\alpha(x, y)$, represent the shifting-phase, respectively.

(2) The measurement process after loading is the same as that before loading. The intensity of the shear speckle interferogram after loading can be expressed as:

\begin{equation}\label{eq2}
\left\{\begin{array}{l}
I_{1}^{\prime}(x, y)=I_{a}+I_{b} \cos [\phi(x, y)+\Delta \varphi(x, y)-3 \alpha(x, y)] \\
I_{2}^{\prime}(x, y)=I_{a}+I_{b} \cos [\phi(x, y)+\Delta \varphi(x, y)-\alpha(x, y)] \\
I_{3}^{\prime}(x, y)=I_{a}+I_{b} \cos [\phi(x, y)+\Delta \varphi(x, y)+\alpha(x, y)] \\
I_{4}^{\prime}(x, y)=I_{a}+I_{b} \cos [\phi(x, y)+\Delta \varphi(x, y)+3 \alpha(x, y)]
\end{array}\right.\,,
\end{equation}

Where  $\Delta \varphi(x, y)$ is the phase change caused by displacement.

(3) The shearography-interferometric phase fringe patterns of the collected sample before and after loading is separated to obtain a series of shear speckle patterns corresponding to the respective wavelengths. The original phase before and after loading can be computed using Carré algorithm.
\begin{equation}\label{eq3}
\bar{\varphi}(x, y)=\arctan \left[\frac{\sqrt{\left[3\left(I_{2}-I_{3}\right)-\left(I_{1}-I_{4}\right)\right]\left[\left(I_{2}-I_{3}\right)+\left(I_{1}-I_{4}\right)\right]}}{\left|\left(I_{2}+I_{3}\right)-\left(I_{1}+I_{4}\right)\right|}\right]\,,
\end{equation}

Where $\bar{\varphi}(x, y)$ are original phase.

According to the Carré algorithm, the original package phase needs to be modified. The modified wrapped phase before and after loading is as follows:
\begin{equation}\label{eq4}
    \varphi(x, y)=\left\{\begin{array}{cc}
\bar{\varphi} & u>0, m>0 \\
\pi-\bar{\varphi} & u>0, m<0 \\
\pi+\bar{\varphi} & u<0, m<0 \\
2 \pi-\bar{\varphi} & u<0, m>0 \\
0 & u=0, m>0 \\
\pi / 2 & u>0, m=0 \\
\pi & u=0, m<0 \\
3 \pi / 2 & u<0, m=0
\end{array}\right.\,,
\end{equation}

Where $\varphi$ are the modified wrapped phase before and after loading, and $u=I_{2}-I_{3}, m=\left(I_{2}+I_{3}\right)-\left(I_{1}+I_{4}\right)$.

The phase difference $\Delta \varphi_{\lambda_{R}}$ corresponding to wavelength ${\lambda_{R}}$ caused by the displacement is as follows:

\begin{equation}\label{eq5}
    \Delta \varphi_{\lambda_{R}}=\varphi_{\lambda_{R}}{ }^{\prime}-\varphi_{\lambda_{R}}\,,
\end{equation}

Where $\varphi_{\lambda_{R}}, \varphi_{\lambda_{R}}{ }^{\prime}$ are the wrapped phase before and after loading corresponding to wavelength $\lambda_{R}$ , respectively. 

Similarly, the wrapped phase difference $\Delta \varphi_{\lambda_{G}}$ corresponding to wavelength  ${\lambda_{G}}$ caused by the displacement can be obtained.

\begin{equation}\label{eq6}
\Delta \varphi_{\lambda_{G}}=\varphi_{\lambda_{G}}{ }^{\prime}-\varphi_{\lambda_{G}}\,,
\end{equation}

Where $\varphi_{\lambda_{G}}, \varphi_{\lambda_{G}}{ }^{\prime}$ are the wrapped phase before and after loading corresponding to wavelength $\lambda_{G}$ , respectively. 

(4) After denoising and unwrapping the wrapped phase caused by displacement, the first derivative of out-of-plane displacement in orthogonal direction is obtained.

\begin{equation}\label{eq7}
\left\{\begin{array}{l}
\frac{\partial w}{\partial x}=\frac{\Delta \varphi_{\lambda_{R}}{ }^{\prime} \lambda_{R}}{4 \pi S_{x}} \\
\frac{\partial w}{\partial y}=\frac{\Delta \varphi_{\lambda_{G}}{ }^{\prime} \lambda_{G}}{4 \pi S_{y}}
\end{array}\right.\,,
\end{equation}

Where $\Delta \varphi_{\lambda_{R}}{ }^{\prime}, \Delta \varphi_{\lambda_{G}}{ }^{\prime}$ are the unwrapped phase corresponding to wavelength $\lambda_{R} \lambda_{G}$ , respectively. $S_{x},S_{y}$ are the shear amount in the orthogonal direction, respectively.

\begin{figure}
\begin{center}
\begin{tabular}{c}
\includegraphics[height=8cm]{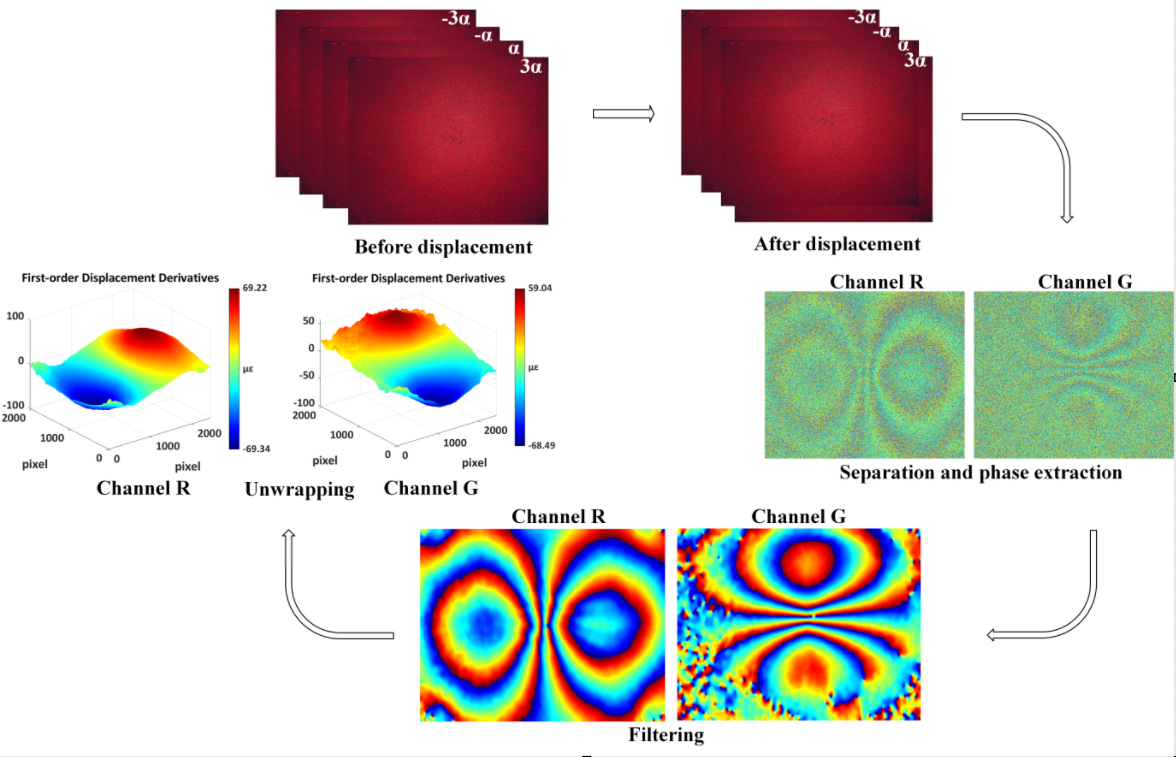}
\end{tabular}
\end{center}
\caption{Procedure of image processing.}
\label{f2}
\end{figure}

Figure~\ref{f2} shows the image processing steps during the experiment. Before and after displacement, four speckle patterns are captured after four equal steps of phase shifting, respectively. The phase difference caused by loading can be calculated using eq. (\ref{eq2})-(\ref{eq4}). The continuous phase can be obtained by filtering and unwrapping.

\begin{figure}
\begin{center}
\begin{tabular}{c}
\includegraphics[width=14cm]{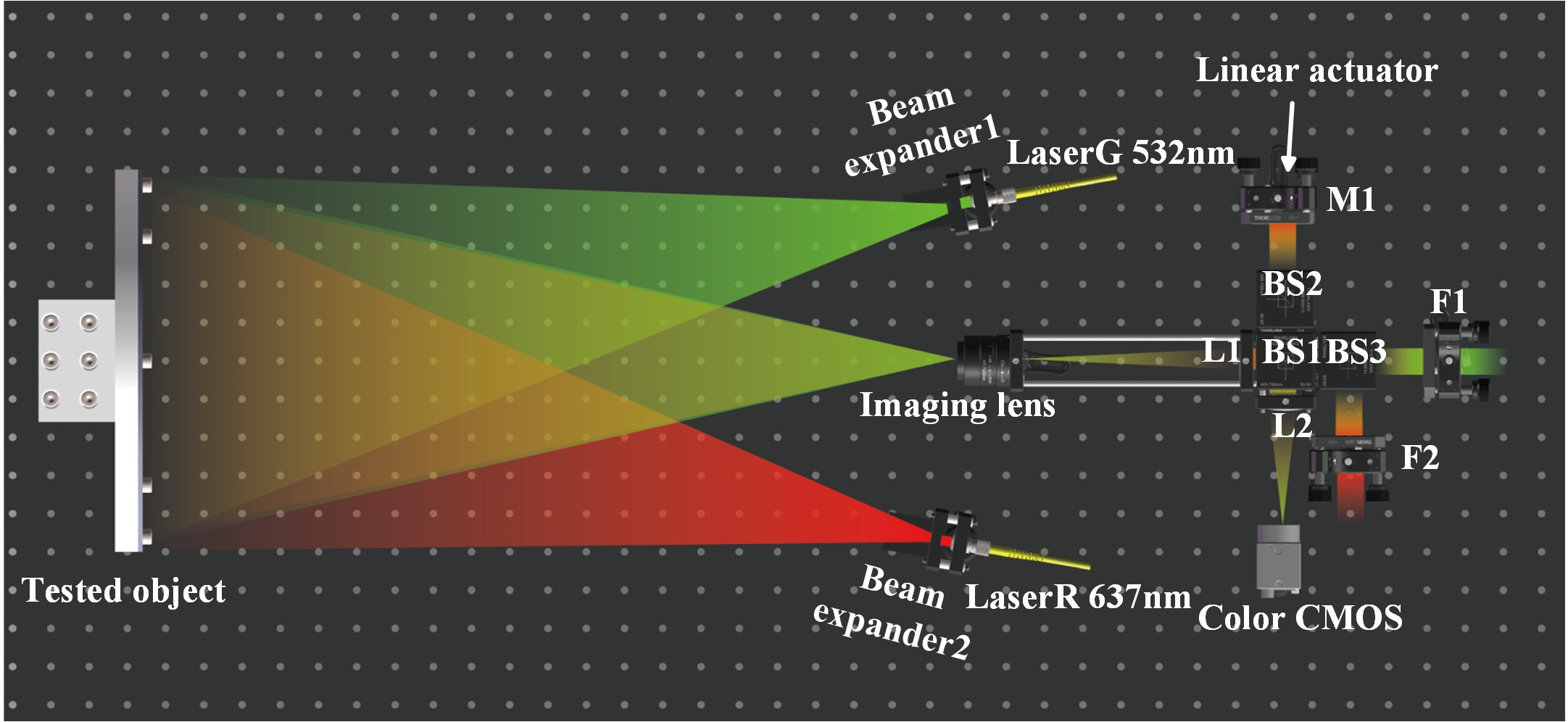}
\end{tabular}
\end{center}
\caption{Three-dimensional drawing of the experiment setup.}
\label{f3}
\end{figure}

\section{Experiment analysis and discussion}

The schematic diagram in Figure~\ref{f1} shows that the experimental system is constructed like Figure~\ref{f3}. A 2048 × 2448-pixel single-chip color CMOS camera with a pixel size of 3.45µm in the horizontal direction and 3.45µm in the vertical direction is used to record the shearography-interferometric phase fringe patterns. A 532nm laser with the maximum output power of 90mw and a 637nm laser with the maximum output power of 110mw were selected to illuminate the test object paraxially. A round metal aluminum plate of diameter 250mm and thickness of 3mm is selected as the test object with central stress loading. The remaining parameters of the experimental setup are consistent with those described in Figure~\ref{f1} .

\begin{figure}
\begin{center}
\begin{tabular}{c}
\includegraphics[width=14cm]{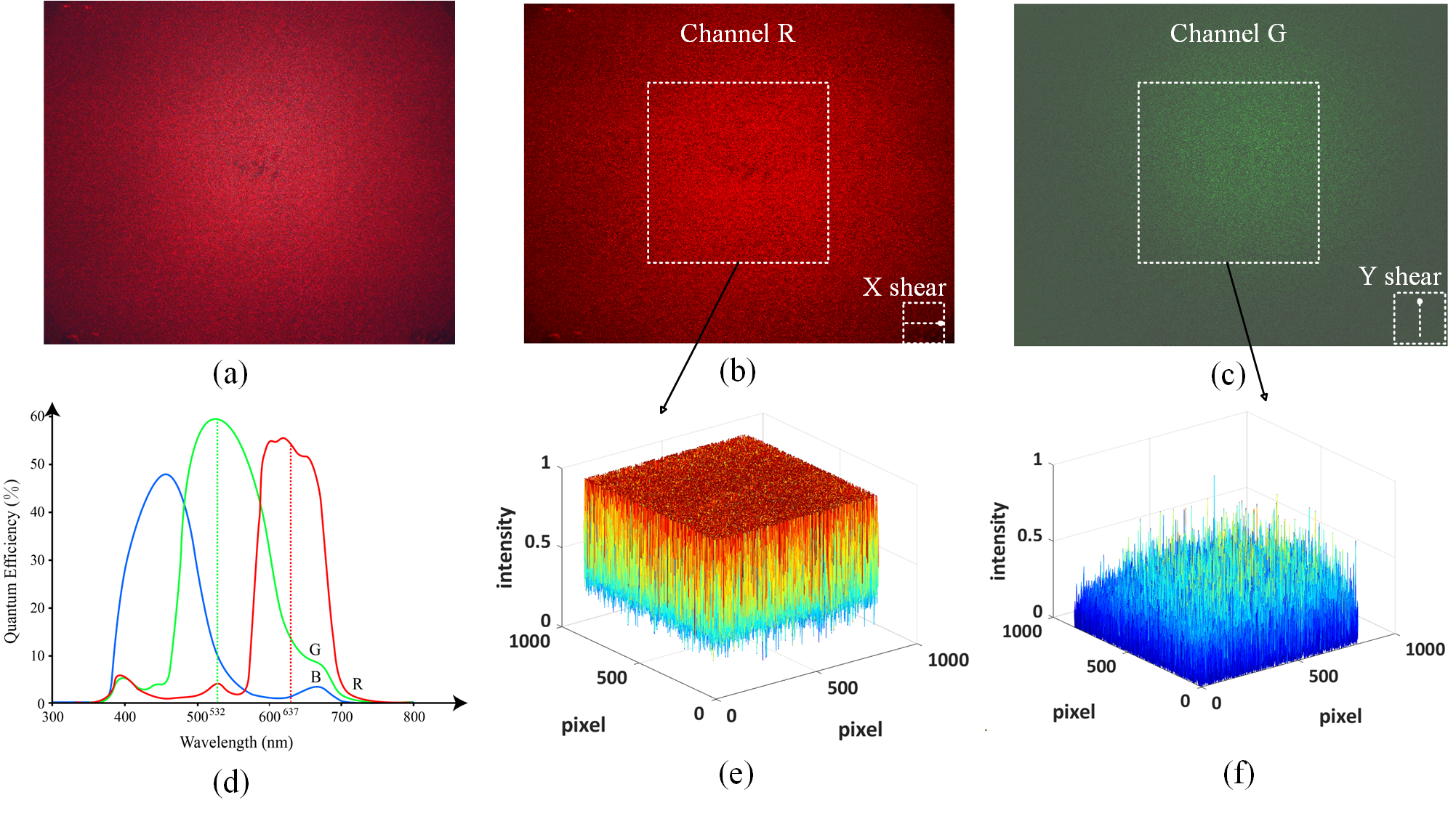}
\end{tabular}
\end{center}
\caption{Analysis results of camera channel extraction when the laser powers of 532nm and 637nm are set to 5mw and 110mw, respectively, and the exposure time of the color camera is set to 50000ms: (a) Original color speckle pattern; (b) X-direction shear results of R-channel extraction; (c) Y-direction shear results of G-channel extraction; (d) Spectral response curve of the single-chip color CMOS; (e) Relative intensity distribution of the central frame selection area of Fig. 4(b); (f) Relative intensity distribution of the central frame selection area of Fig. 4(c).}
\label{f4}
\end{figure}

The sensitivity curve of the color camera used in the experimental device is shown in Figure~\ref{f4}(d). Since the camera is a single-chip color CMOS, the quantum efficiency curves of the three RGB channels in Figure~\ref{f4}(d) overlap, which may easily lead to aliasing of the images after channel separation. In the experiment, it is found that the power of the 532nm laser is too high, and there will be a strong signal response in the R channel besides G channel, which will affect the information extraction of the R channel. By controlling the power of the 532nm and 637nm lasers and the exposure time of the camera, the influence of channel aliasing is reduced and a better image channel extraction is achieved. According to the experimental results, for the lasers we use, the laser powers of 532nm and 637nm are set to 5mw and 110mw, respectively, and the exposure time of the color camera is set to 50000$\mu s$. At this time, the R and G channels information extracted has the least influence on each other. Figure~\ref{f4}(a) shows a speckle pattern under this setting. The corresponding speckle patterns after channel R and G extraction are shown in Figure~\ref{f4}(b) and (c), respectively. According to dislocation directions of the screw with the corresponding color signals in the lower right corner of Figure~\ref{f4}(b) and (c), it is judged that the corresponding shear directions of R and G channels are x and y, respectively. The intensity distribution in the central frame selection area of Figure~\ref{f4}(b) and (c) corresponds to Figure~\ref{f4}(e) and (f), respectively. According to Figure~\ref{f4}(e) and (f), the relative average intensity corresponding to the R and G channels is 0.6981±0.1907 and 0.1271±0.0849, respectively. 

\begin{figure}
\begin{center}
\begin{tabular}{c}
\includegraphics[width=14cm]{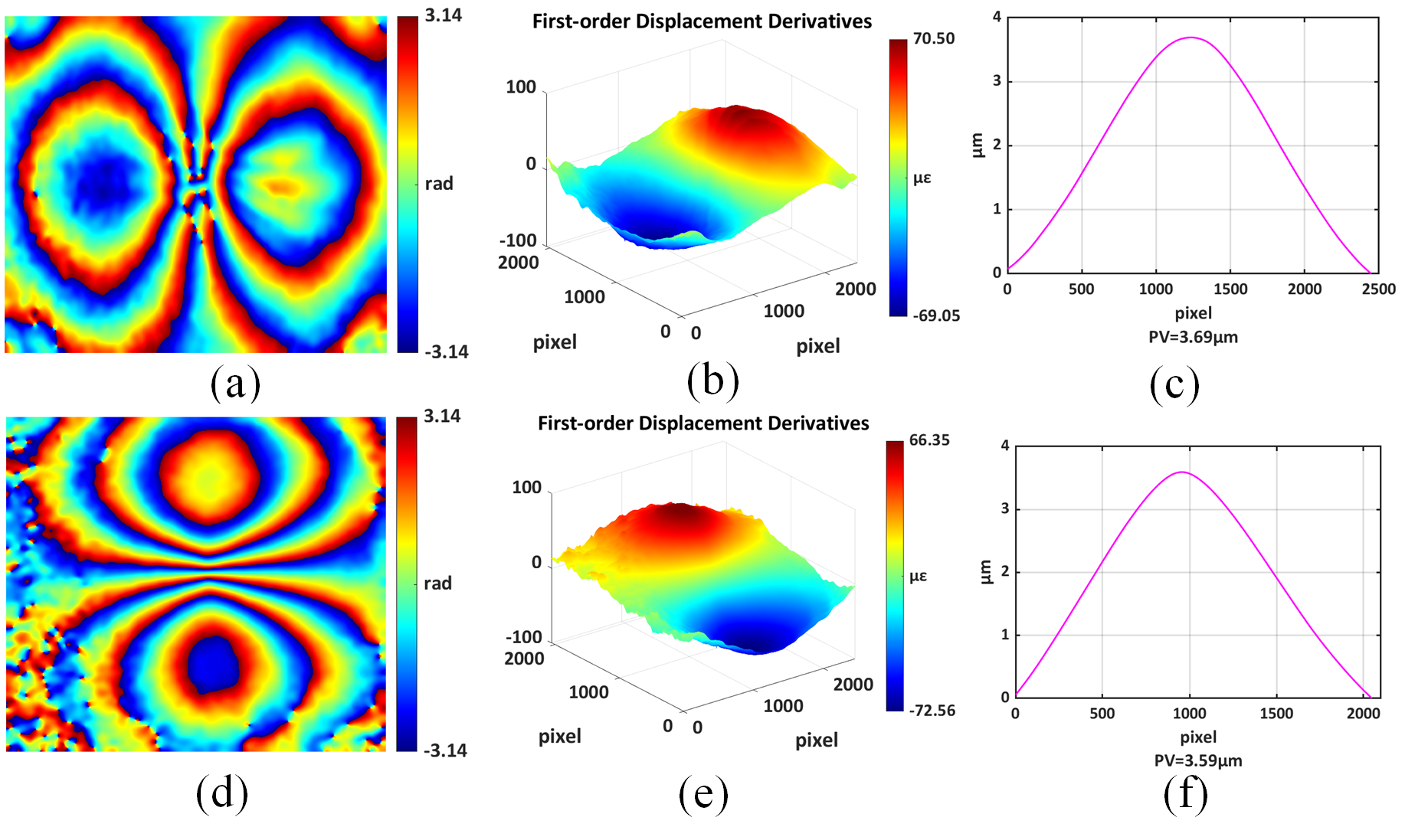}
\end{tabular}
\end{center}
\caption 
{ \label{f5}
Experiment results when the laser powers of 532nm and 637nm are set to 5mw and 110mw, respectively, and the exposure time of the color camera is set to 50000$\mu s$: X shear direction: (a) Wrapped phase pattern; (b) The first-order displacement derivative after unwrapping; (c) Displacement and PV along the central axis in the x direction. Y shear direction: (d) Wrapped phase pattern; (e) The first-order displacement derivative after unwrapping; (f) Displacement and PV along the central axis in the y direction. } 
\end{figure} 

Experiment results under above settings are shown in Figure~\ref{f5}. The field of the experimental system is 0.167mm/pixel. The shear amounts of x and y direction are 9.67mm and 10.33mm, respectively. Figure~\ref{f5}(a) and (d) are wrapped phase pattern along x and y shear direction after channel separation, phase extraction and filtering, respectively. The wrapped phase information of the two shear directions is relatively complete and has no mutual influence. The first-order displacement derivative after unwrapping along x and y shear direction are shown in Figure~\ref{f5}(b) and (e), respectively. Variation range of the first-order displacement derivative along x and y shear direction are [-69.05 $\mu \varepsilon$, 70.50$\mu \varepsilon$] and [-72.56$\mu \varepsilon$, 66.35$\mu \varepsilon$], respectively. Displacement and PVs along the central axis in the x and y shear directions are shown in Figure~\ref{f5}(c) and (f), respectively. The PVs of displacement along x and y shear direction are 3.69µm and 3.59µm, respectively. The relative error is 2.71\%.

\begin{figure}
\begin{center}
\begin{tabular}{c}
\includegraphics[width=14cm]{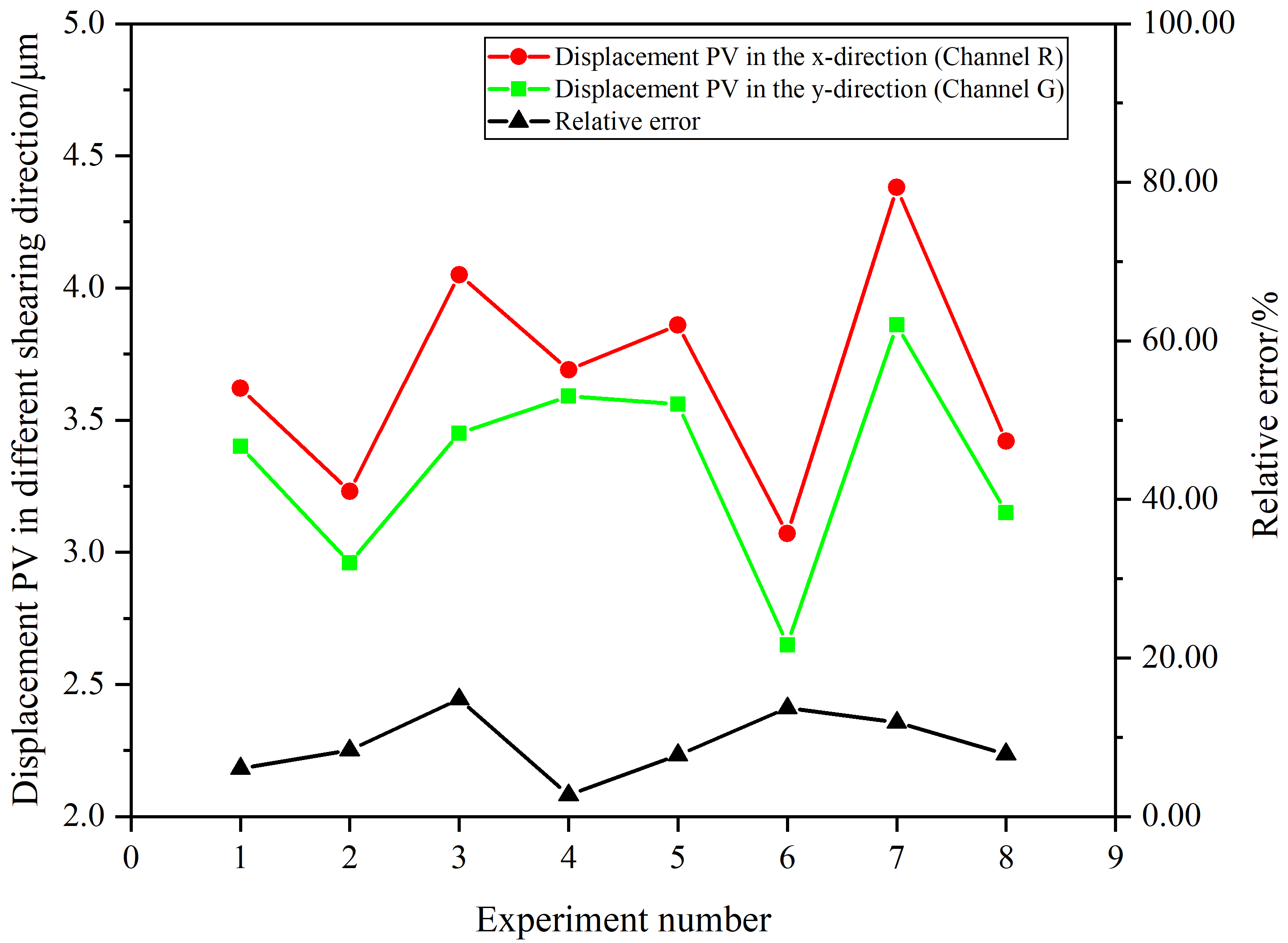}
\end{tabular}
\end{center}
\caption 
{ \label{f6}
Displacement PV and relative error in the x (R channel),y (G channel) shear direction in multiple experiments. } 
\end{figure} 

Figure~\ref{f6} shows the displacement PV and relative error in x and y shearing directions in multiple experiments. The experimental results show that the overall displacement integral error in the x and y directions is 2.7\%-14.8\%. When the loading is relatively small and under 4µm, the relative error of the displacement integral is small. The deformation PV in the x shearing direction is always larger than that in the y direction, which is related to the fact that the green light power is smaller and the uneven illumination leads to poor fringes in the y direction.

\section{Conclusion}

This paper proposed a novel temporal phase-shift digital shearography system including dual lasers with wavelengths of 532nm and 637nm, three splitter prism structure, two dichroic mirrors with different response wavelength, and the color CMOS to measure simultaneously first order displacement derivative in orthogonal shear directions. The lasers with wavelength of 532nm and 637nm set different the emission power to avoid color camera R, G channel signal overlap and realize the separation of speckle information in orthogonal directions. The experiment completely measured the first-order displacement derivative information in the orthogonal direction, and verified the reliability of the system through the displacement PV in two directions. The proposed temporal phase-shift digital shearography system can be applied to the measurement of first-order displacement derivatives in orthogonal directions such as composite materials, multi-directional defect detection, and improve defect detection capabilities. In further research, it is necessary to improve the fringe quality by adjusting and improving the green light power, the gain of color CMOS and exposure time, so as to reduce the error and improve the measurement accuracy.

\section{Disclosures}
The authors declare no conflicts of interest.

\section{Code, Data, and Materials Availability}
The data supporting this study’s findings are available upon reasonable request from the authors.

\section{Acknowledgments}
This research was funded by the National Natural Science Foundation of China (No.51535003 and No.62205316) and Local Science and Technology Development Fund Projects Guided by the Central Government, China (No.2022ZYDF065).


\bibliography{report}   
\bibliographystyle{spiejour}   


\vspace{2ex}\noindent\textbf{Yinhui Guo} graduated from Shandong University with a bachelor's degree. He is now studying in Laser Fusion Research Center , China Academy of Engineering Physics and engaged in the research of digital shearography.

\vspace{1ex}
\noindent Biographies and photographs of the other authors are not available.

\listoffigures

\end{spacing}
\end{document}